%% file: sdess.tex
\documentclass[conference]{IEEEtran}
\ifCLASSINFOpdf
\else
\fi
\usepackage[cmex10]{amsmath}
\usepackage{url}

\usepackage{mathptmx}
\usepackage{amsmath}
\usepackage{mathptmx}

\usepackage{times}
\usepackage{xcolor}
\usepackage{lastpage}
\usepackage{amsmath}
\usepackage[utf8]{inputenc}
\usepackage{epsfig,endnotes,url,color}
\usepackage{array}
\usepackage{multirow}
\usepackage{xspace}
\usepackage{verbatim}
\usepackage{wrapfig}
\usepackage{mathptmx}
\usepackage[english]{babel}
\usepackage{graphicx}
\graphicspath{{./}{figs/}}
\usepackage{csvsimple}
\usepackage{epstopdf}
\usepackage{subcaption}
\usepackage{enumitem}
\usepackage{datetime}
\usepackage{url}
\usepackage{cleveref}
\usepackage{amssymb}
\usepackage{tabularx}
\usepackage{mathptmx}
\usepackage[font=small,skip=0pt]{caption}

\RequirePackage{snapshot}

\newcommand{\projecttitle}{{\scshape Authorized Analytics}\xspace}

\begin{document}
%
\title{Mobile Privacy-Preserving Crowdsourced Data Collection in the Smart City}

\author{
\IEEEauthorblockN{Joshua Joy\IEEEauthorrefmark{1},Ciaran McGoldrick\IEEEauthorrefmark{2},Mario Gerla\IEEEauthorrefmark{3}}
\IEEEauthorblockA{UCLA\\Email:\IEEEauthorrefmark{1}jjoy@cs.ucla.edu,\IEEEauthorrefmark{2}ciaran@cs.ucla.edu,\IEEEauthorrefmark{2}gerla@cs.ucla.edu}
}


%


\maketitle

%
\IEEEpeerreviewmaketitle

\input{abstract}
\input{introduction}

\input{background}
\input{related}

\input{goals}

\input{architecture}

\input{evaluation}

\input{conclusion}

\input{acknowledgements}

{\footnotesize
\bibliographystyle{acm}
\bibliography{sdess,main}
}

\end{document}

%% file: abstract.tex
\begin{abstract}

Smart cities rely on dynamic and real-time data to enable smart urban applications such as intelligent transport and epidemics detection. However, the streaming of big data from IoT devices, especially from mobile platforms like pedestrians and cars, raises significant privacy concerns.

Future autonomous vehicles will generate, collect and consume significant volumes of data to be utilized in delivering safe and efficient transportation solutions. The sensed data will, inherently, contain personally identifiable and attributable information - both external (other vehicles, environmental) and internal (driver, passengers, devices).

The autonomous vehicles are connected to the infrastructure cloud (e.g., Amazon), the edge cloud, and also the mobile cloud (vehicle to vehicle). Clearly these different entities must co-operate and interoperate in a timely fashion when routing and transferring the highly dynamic data.  In order to maximise the availability and utility of the sensed data, stakeholders must have confidence that the data they transmit, receive, aggregate and reason on is appropriately secured and protected throughout.   There are many different metaphors for providing end-to-end security for data exchanges, but they commonly require a management and control sidechannel.

This work proposes a scalable smart city privacy-preserving architecture named \projecttitle that enables each node (e.g. vehicle) to divulge (contextually) local privatised data.  \projecttitle is shown to scale gracefully to IoT scope deployments. 

\end{abstract}

%% file: introduction.tex
\section{Introduction}

\begin{figure}
    \centering
    \includegraphics[width=1\columnwidth]{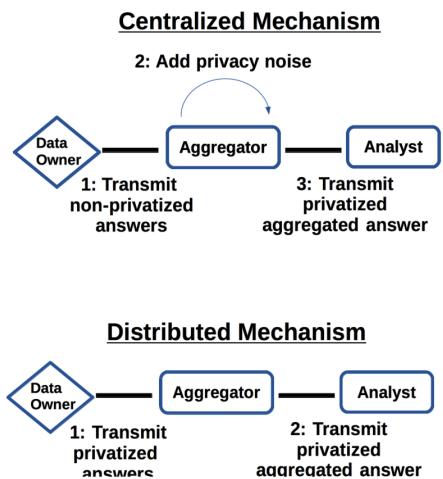}
    \caption{Illustrates that the data owner has control and consent over the privatization release as opposed to the centralized mechanism which requires strong trust assumptions regarding the aggregator adding differentially private noise.}
    \label{fig:privatization}
\end{figure}

\begin{table*}[t]
\centering

\begin{tabular}{l|l|l|l|l|l|l|l|l}
\hline
\textbf{}                                     & \textbf{\begin{tabular}[c]{@{}l@{}}Output\\ Pertubation\end{tabular}} & \textbf{\begin{tabular}[c]{@{}l@{}}Input\\ Pertubation\end{tabular}} & \textbf{\begin{tabular}[c]{@{}l@{}}Trusted\\ Aggregator\end{tabular}} & \textbf{\begin{tabular}[c]{@{}l@{}}Untrusted\\ Aggregator\end{tabular}} & \textbf{\begin{tabular}[c]{@{}l@{}}Pollution\\ Protection\end{tabular}} & \textbf{Anonymity} & \textbf{\begin{tabular}[c]{@{}l@{}}Stream\\ Analytics\end{tabular}} & \textbf{\begin{tabular}[c]{@{}l@{}}Sparse\\ Datasets\end{tabular}} \\ \hline
PINQ\cite{DBLP:conf/nsdi/ChenRFG12}           & X                                                                            &                                                                                  & X                                                                     &                                                                         &                                                                         &                    &                                                                     &                                                                    \\ \hline
PDDP\cite{DBLP:conf/nsdi/ChenRFG12}           &                                                                              & X                                                                                & X                                                                     &                                                                         & X                                                                       &                    &                                                                     & X                                                                  \\ \hline
SplitX\cite{DBLP:conf/sigcomm/ChenAF13}       &                                                                              & X                                                                                & X                                                                     &                                                                         & X                                                                       & X                  &                                                                     & X                                                                  \\ \hline
Rappor\cite{DBLP:conf/ccs/ErlingssonPK14}     &                                                                              & X                                                                                &                                                                       & X                                                                       &                                                                         &                    &                                                                     &                                                                    \\ \hline
PCDH-LU\cite{DBLP:conf/pet/ChanLSX12}         &                                                                              & X                                                                                &                                                                       & X                                                                       &                                                                         &                    & X                                                                   &                                                                    \\ \hline
Safe Zones\cite{DBLP:conf/ndss/FriedmanSKS14} & X                                                                            &                                                                                  & X                                                                     &                                                                         &                                                                         &                    & X                                                                   &                                                                    \\ \hline
\projecttitle                                 &                                                                              & X                                                                                &                                                                       & X                                                                       & X                                                                       & X                  & X                                                                   & X                                                                  \\ \hline
\end{tabular}

\caption{Privacy-Preserving Stream Analytics Design Space}
\label{tab:designspace}
\vspace{-2.0em}
\end{table*}

Researchers are becoming increasingly interested in studying smart city behaviors, like pedestrians, drivers and traffic, city resources (e.g., energy) and city environment (e.g., pollution, noise). These studies are commonly based on Open Shared Data made available by several Smart City testbeds around the country. To this end, Open Data Science enables researchers to collect the data, analyze and process it with Data Mining and Machine Learning techniques and create accurate models that allow them to credibly validate smart city design methodologies. These systems enable the collection of data from sensors, cameras embedded in the "smart city" (e.g., smart building, smart transport, smart instrumented crowds) which can be used to derive models of behavior, predict trends, optimize system management and detect the onset of attacks.

There is now an increasing demand that research addressing these challenges be performed in more realistic environments. In other words, researchers will need to deploy their technologies in real vehicles, in real roads and cities (or, at smaller-scale, on-campus roads used for general purposes), to demonstrate that they are not mere simulation and pilot testbed toys and do scale to urban dimensions.  As smart city research and systems require testing and validation in such uncontrolled environments, considerable attention must be paid to the validity of the experiments and the integrity and privacy of the data gathered through them.  Since the experiments must be performed on massive scale in public places, it is prudent to anticipate malicious agents who either wish to make illicit use of the data gathered or seek to inject false data.

As this research will have significant impact on the economy and safety of smart environments, the security challenges to realistic in-the-field experiments carried out in the smart city must be addressed.  Rather than requiring each researcher working in the area to start afresh on securing his experimental infrastructure, there should be a common ``privacy" infrastructure that all researchers can adopt.  This approach has several advantages, including allowing researchers to concentrate on the issues they care about, rather than addressing issues of cybersecurity that, while important, are not relevant to the research questions they seek to answer. A shared cybersecure infrastructure can assure that security holds for a representative set of technologies and approaches, rather than requiring each project to perform a separate security evaluation.

In this work named \projecttitle, the vehicle example will be used to illustrate secure and resilient IOT data gathering, as vehicles are extremely rich sensor platforms, run mature applications and touch many critical smart city issues, from citizens’ welfare to energy and environment. In dealing with vehicles and their associated data, a central issue will be privacy e.g. of location data. Using techniques that go under the generic name of Differential Privacy~\cite{DBLP:conf/icalp/Dwork06,DBLP:conf/tcc/DworkMNS06,DBLP:conf/eurocrypt/DworkKMMN06}, a vehicle that wishes to remain anonymous will obfuscate(fuzz) its position so that subsequent observations cannot reveal the exact origin and destination of the trip, yet permits analyzers to extract meaningful analysis from the aggregate data. The methodology is not restricted to vehicles, and can be extended to many other massively distributed urban computing  observation platforms, from intelligent homes and smart energy grids to pedestrians and drones, etc.
 
The goal of this work is to develop an integrated architecture that allows resilient and secure gathering of Internet Of Things (IOT) data in the face of impairments and attacks that can originate at different layers of the network protocol stack. Moreover, privacy and confidentiality can be efficiently assured as required by the application.  The role of the urban infrastructure (in the form of mobile cloud and edge cloud) will be elaborated. The architecture is realised and demonstrated in an important emerging IOT scenario, the autonomous vehicle.

%% file: background.tex
\section{Background}

Table~\ref{tab:designspace} gives an overview of various privacy-preserving properties of related work and our proposed target model named \projecttitle. We now explain the desired properties and motivate \projecttitle.

Perturbation techniques are classified into two basic categories: input perturbation and output perturbation \cite{DBLP:conf/tcc/DworkMNS06}.Input perturbation techniques are those where the underlying data are randomly modified, and answers to questions are computed using the modified data resulting in noisy answers. Output perturbation techniques are those where (correct) answers to queries are computed exactly from the real data, but noisy versions of these are reported.

Figure~\ref{fig:privatization} demonstrates the difference between a trusted aggregator and untrusted aggregator.

A trusted aggregator sits between the data owners and the analysts, verifies and may modify the query, and returns the aggregate sum to the analysts \cite{DBLP:conf/icalp/Dwork06,DBLP:conf/tcc/DworkMNS06}. A trusted aggregator must blindly add noise and add the exact amount of noise (no more no less). If not enough differential private noise is added, the differential privacy guarantees do not hold breaking the privacy preserving guarantees. If too much diffferential private noise is added, the answers will be too noisy and not useful.

An untrusted aggregator computes and publishes some statistics based on the privatized answers received from the data owners. The data may live indefinitely as it is the data owner, rather than the aggregator, which adds the differential private noise. Thus, even if the aggregator were to be breached, no privacy violations would occur.

It's important that a single malicious adversary is unable to submit a single answer (e.g., very large answer) that substantially distorts the aggregate sum. Safeguards against this attack is commonly referred to as pollution protection.

Data owners also desire to remain anonymous within the set of all other participating data owners (anonymity set). The anonymity guarantee ensures that private answers submitted are unable to be linked back to the specific data owner by either the aggregator or a malicious adversary performing traffic analysis.

To ensure timely results support for stream analytics is desirable. This ensures that data is processed continuously and released on the order of minutes.

Finally, it is important that the accuracy of the results is able to scale from small datasets (e.g., small IoT deployments) to large scale IoT systems. This means that both sparse datasets and heavy hitter algorithms should be supported by the system.

Thus, it can be seen that \projecttitle should include the desired properties of input pertubation, untrusted aggregator, pollution protection, anonymity, stream analytics, and sparse datasets. This ensures protection against data breaches at the aggregator, graceful scalability for large scale IoT deployments, and high accuracy.

%% file: related.tex
\section{Related}

Privacy-preserving analytics has been an active area of research in recent years~\cite{DBLP:conf/ndss/FriedmanSKS14,DBLP:conf/ccs/HardtN12,DBLP:conf/pet/ChanLSX12,DBLP:conf/nsdi/ChenRFG12,DBLP:conf/stoc/DworkNPR10,DBLP:conf/ccs/ErlingssonPK14,DBLP:conf/sigmod/RastogiN10,DBLP:conf/ndss/ShiCRCS11}.

In distributed data analytics systems~\cite{DBLP:conf/ccs/HardtN12,DBLP:conf/nsdi/ChenRFG12,DBLP:conf/sigcomm/ChenAF13}, data owners maintain their personal data locally, and have exclusive control over their own data.  These systems enable analysts to analyze data owners' locally maintained data in a distributed manner under the differential privacy guarantees.  However, the systems~\cite{DBLP:conf/nsdi/ChenRFG12} rely on expensive public-key crypto systems which do not scale well in a large-scale IoT environment and are not suitable for stream analytics.  Subsequent system~\cite{DBLP:conf/ccs/HardtN12} addresses the scaling issue by avoiding using expensive crypto systems.  It, however, suffers from the pollution attack where even a single malicious data owner can substantially distort the query result without detection.  The recent system~\cite{DBLP:conf/sigcomm/ChenAF13} defeats the pollution attack, however, it involves some level of synchronization among system components preventing its usage for stream analytics.  Nevertheless, all these systems require trusted servers, and deal with only the ``one-shot'' queries whereby data owners' data is assumed to be static and the aggregate result remains unchanged during the course of the query execution.

Data analysts care about timeliness as it is common that user
data is constantly changing. It is crucial that analysts can analyze user data continuously.  Dwork et al.~\cite{DBLP:conf/stoc/DworkNPR10} proposed one of the first proposals to make the transition from differentially private static analytics to stream analytics.  This system introduces the concept of pan-privacy. Data owners reveal their data relying on the pan-private mechanism to defend against intrusions. However, this system updates the query result only after data owners' data changes significantly, and does not support stream analytics over an unlimited time period.

Chan et al.~\cite{DBLP:conf/pet/ChanLSX12} operate with an untrusted aggregator, and support differentially private continual monitoring of multiple distributed streams over a sliding window. Similarly, Friedman et al.~\cite{DBLP:conf/ndss/FriedmanSKS14} propose a concept of safe zones, and enable the monitoring of arbitrary functions over statistics derived from distributed data streams in a differentially private manner.  However, these systems are tailored for heavy-hitter monitoring only (i.e., they can only report on a fraction of the data), and require some form of synchronization which limits their scalability.

Recently, RAPPOR~\cite{DBLP:conf/ccs/ErlingssonPK14} enables analysts to collect various types of statistics from a large number of data owners' devices in such a way that provides the differential privacy guarantees for individual data owners. However, RAPPOR is designed for heavy-hitter collection, and does not deal with the situation where data owners' answers to the same query are changing over time.  Therefore, RAPPOR does not fit well with the IoT stream analytics.

%% file: goals.tex
\section{Goals}

\begin{figure}[ht!]
    \includegraphics[width=1\columnwidth]{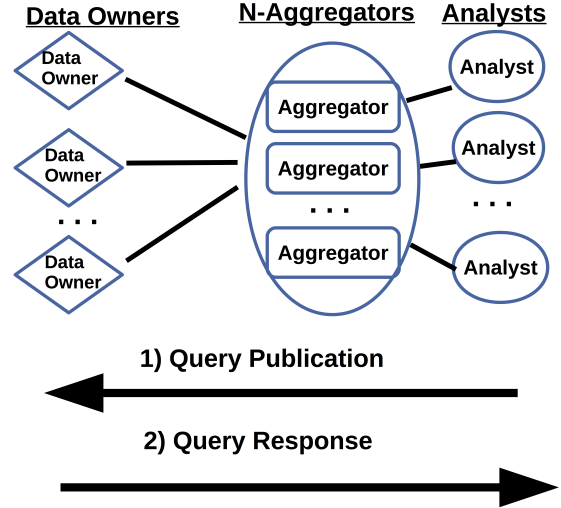}
    \caption{System architecture. Data owners privately and anonymously respond in real-time to the analysts? queries. The privatized responses are continuously aggregated into small batches and then transmitted to each corresponding analyst.}
\vspace{-2.0em}
    \label{fig:architecture}
\end{figure}

We propose to enable scalable and privacy-preserving data analytics for personal identifiable information (PII) that protects individuals' privacy yet enables analysts to leverage available data to improve urban scenarios, healthcare, traffic regulation (congestion avoidance, accident prevention), and crime control (locating lost/stolen vehicles). Vehicular networks are targeted as a challenging exemplar. Our vision is to tap into underutilized real time data - both vehicular (personal (seatbelt use, car seat use, etc) and public (pollution, traffic condition, etc.)).This data  is continuously generated and (generally) cannot be stored in central cloud infrastructures for two reasons -- users' privacy concerns, and the volume and timeliness requirements from the generated data.

\textbf{Trust Assumptions.} Analysts are potentially malicious. Analysts may attempt to de-anonymize data owners, link data owners' query requests and answers, or remove the differential private ($DP$) noise added to query answers. Data owners are also potentially malicious.  They may generate false or illegitimate answers in an attempt to distort the query results learned by analysts.  We assume that there are no Sybil data owners (i.e., data owner create multiple false identities) and delegate to other procedures (e.g., vehicular network security based on public-key infrastructure) for the elimination of Sybil nodes/identities. 

Finally, the aggregators are also potentially malicious.  Alternatively, they may malfunction and not follow the specified protocols faithfully.  This may invalidate the query results, but will not be able to violate the privacy of individual users due to the utilization of input perturbation. Prior work must synchronize the addition of differentially private noise~\cite{DBLP:conf/sigcomm/ChenAF13} and must be trusted to safeguard the unprivatized data and properly release the privatized answers faithfully~\cite{DBLP:conf/nsdi/ChenRFG12,DBLP:conf/sigcomm/ChenAF13}, leaving these prior systems vulnerable to data breaches.

\textbf{Performance Properties}
Current state-of-the-art systems provide differential privacy guarantees. However, they fall short in applications with ``one-shot'' queries. One-shot query systems have strong and unrealistic assumptions -- user data is assumed to be static and the aggregate result is assumed unchanged spanning hours and days. Recently, data analysts have begun to care about timeliness, especially when user data is constantly changing. This is particularly true in mobile networks where the mobiles (eg. vehicles) may upload new data every minute. 

The privacy-preserving software should scale well both on the data owners' mobile devices as well as on the server system components (i.e., the aggregator). The data owner-side operations should not incur much overhead to support even the most resource-constrained devices, e.g., mobile phones and low-capacity sensors.  The server-side operations should be efficient enough to continuously support a large data owner base with millions of data owners on the order of minutes; additionally, the servers should tolerate data owner churn as data owners may be intermittently disconnected. 

\textbf{Privacy Properties.}
Data owners' personal data resides on their own devices.  Data owners have exclusive control over their own data, and are able to determine whether or not to answer queries over their own data.  

The system should satisfy the local differential privacy mechanism as well as provide anonymous guarantees robust to traffic analysis. The local differential privacy mechanism ensures that each data owners response should be independently generated such that a single data owner's response does not rely on the addition of differentially private noise from another data owner. Thus, even if an active attack were to be made against a single data owner or the aggregator, the differential privacy guarantee holds true. The anonymity property ensures that any answer submitted is unable to be linked to a single data owner.

%% file: architecture.tex
\section{Architecture}

\subsection{System Model}

The system under study is a mobile IoT system that collects data samples from the environment (external) and from the data owner (internal). Personal data is stored in each device and privately  released upon the device answering a query from analysts. Privacy is a concern, but so is real time response and quality of information.  Privacy, latency and utility are the main focus of the study. The system model consists of three components - IoT devices, aggregators, and analysts as seen in Figure~\ref{fig:architecture}.

The mobile IoT devices considered in our study are vehicles, because vehicles are the mobile agent(s) with the richest assortment of on board sensors, with  efficient communications to peers and to infrastructure and with (effectively) unlimited power supply. Vehicles can communicate either with other moving vehicles or with fixed network nodes placed alongside the road, commonly referred to as road-side units (RSUs). The typical use of an RSU is to provide moving vehicles with access to an infrastructure network, as well as infrastructure-based services. 

While there have been various prior work which proposed privacy mechanisms~\cite{Sweene02,DBLP:conf/icde/MachanavajjhalaGKV06,DBLP:conf/tcc/DworkMNS06}, we choose differential privacy ($DP$) as differential privacy has gained broad acceptance. Recent data analytics systems achieve $DP$ guarantees in a distributed setting~\cite{DBLP:conf/eurocrypt/DworkKMMN06,DBLP:conf/ccs/HardtN12,DBLP:conf/nsdi/ChenRFG12,DBLP:conf/sigcomm/ChenAF13} where each data owner holds their own personal data, rather than in a centralized database. Specifically, in these systems, the query answers are first generated at each data owner's own device independently, and then the $DP$ noise is added to the aggregate answer (either collaboratively by the data owners or by a centralized aggregator). However, existing distributed systems require strong trust assumptions regarding the aggregation mechanism and require expensive zero-knowledge proofs to defend against pollution attacks (e.g., where one data owner can distort the aggregate sum with a single answer). 

We are interested in Distributed $DP$ (local differential privacy) mechanisms whereby each data owner holds their own data and is independently responsible for adding differentially private noise. By satisfying local $DP$, strong trust assumptions are removed, pollution attacks are thwarted, and data owners are guaranteed $DP$ regardless of the amount of noise added by other data owners or system components.

\begin{enumerate}
\item \textbf{Analysts.} An analyst formulates queries, distributes the queries to data owners via the aggregator, and receives the query results from the aggregator. Multiple analysts can operate in the system simultaneously. Each query is signed by the analyst. Examples of an analyst may be the \textit{Los Angeles Department of Transportation}, \textit{Center for Disease Control}, or \textit{National Institutes of Health}.

\item \textbf{Data Owners.} A data owner is a user device or agent under the user's control.  It stores the continuously-generated personal user data locally.  Each data owner device requests queries from specific analysts, formulates answers to these queries using the data owners' local data, and responds with privatized and anonymized answers. All the query requests and answers are anonymized and are TLS transmitted via aggregators.

\item \textbf{Aggregator.} The data owners transmit their privatized and anonymized answers to the aggregator via TLS. The aggregator interfaces with analysts, aggregates the small batches of responses, and provides the privacy-preserving stream analytics service.  
\end{enumerate}

Additionally, it should be noted that the anonymity scheme still provides a form of accountability that prevents Sybil attacks. All anonymous users will be authenticated, for example using PKI (managed by the analysts for example), such that only authenticated data owners are allowed to upload data. All data owners form an anonymity set, such that a single data owner is anonymous with the set of all other uploading data owners. The proposed privacy preserving solutions are an integral part of the overall architecture and offer greatly reduced processing overhead and latency.

\noindent \textit{Query Formulation.} An example query may be ``what is the distribution of real-time speed across the Los Angeles metro area?". Here, the analyst can define 22 indexes on speed: `0', `1$\sim$10', `11$\sim$20', $\cdots$,  `181$\sim$190', `191$\sim$200', and `\textgreater200'.  If a vehicle is moving at 15 mph, the data owner on that vehicle answers `1' for the third index, and `0' for all others.

Thus, each query specifies a few fields as follows:
\begin{equation}
\label{eqn:query}
Query := \langle Query_{ID}, Analyst_{ID}, Sensors_{1 \to n}, p, q, epoch, T_{end} \rangle
\end{equation}

Each query is uniquely identified by the query $ID$ and analyst $ID$. The query is formulated as an $n$-dimensional bit vector over the IoT sensors.

$Sensors_{1 \to n}$ denotes the $n$ sensors to be monitored. Upon querying each sensor, the retrieved sensor value is then matched against the formulated query indexes. Each data owner's answer is in the form of a `1' or `0' per index, depending on whether or not the answer falls within that range.  Different ranges should not overlap. 

$p$ and $q$ denote the two coin flipping probabilities used in the randomized response mechanism later at data owners.  They describe the probabilities of the first coin and the second coin coming up heads. The analyst requests a certain utility (express by p,q) and the data owners may refuse if too risky.

$epoch$ denotes the execution interval of the query, i.e., how often the query needs to be executed at a data owner. $T_{end}$ denotes the query end time, i.e., when answers to the query will no longer be accepted.

\noindent \textit{Sanity Check.} The data owner's device checks all the queries received from analysts and discards any query if its privacy cost $\epsilon$ (computed based on the query's $p$ and $q$) exceeds a certain configurable threshold.  The data owner is also able to restrict query execution based on a personal sensor blacklist. Analyst reputation may also be used to blacklist certain analyst queries from being executed.

\subsection{Privacy Mechanism}
\label{sec:randomizedresponse}

Randomized response~\cite{warner1965randomized} was originally created by social scientists as a privacy mechanism to allow individuals to answer sensitive questions without providing truthful answers all the time yet still allow analysts to collect meaningful statistical user data.  Utilizing the randomized response method, data owners locally randomize their truthful answers to analysts' sensitive queries, and respond only with the privatized (locally randomized) answers to the analysts. Randomized response mechanism also satisfies the differential privacy guarantee as well as provides the optimal sample complexity for local differential privacy mechanisms~\cite{DBLP:conf/nips/DuchiWJ13}.

There are many different randomized response mechanisms in the literature.  In this section, we present only the mechanism described in~\cite{fox1986randomized} because it offers a superior balance between the utility and the privacy guarantee of randomized responses, as compared to other mechanisms~\cite{warner1965randomized,kuk1990asking,greenberg1969unrelated}.

\paragraph{Mechanism Description} Suppose each data owner has two independent coins.  With coin flipping, the first coin comes up heads with probability $p$, and the second coin comes up heads with probability $q$.  Without loss of generality, in this paper, heads is represented as ``yes'' (i.e., 1), and tails is represented as ``no'' (i.e., 0).

The data owner flips the first coin. If it comes up heads, the data owner responds with the truthful answer; otherwise, the data owner flips the second coin and reports the result of this second coin flipping.  Eventually, if there are $N$ randomized answers where $Y_R$ of them are ``yes'', then the estimated number of the original truthful ``yes'' answers (i.e., without randomization), $Y_E$, can be computed as:
\begin{equation}
\label{eqn:yo}
Y_E = \frac{Y_R - (1 - p) \times q \times N}{p}
\end{equation}

The intuition behind randomized response is that it provides ``plausible deniability'', i.e., any truthful answer can produce a response either ``yes'' or ``no'', and data owners retain strong deniability for any answers they respond. If the first coin always comes up heads, there is high utility yet no privacy. Conversely, if the first coin is always tails, there is low utility though strong privacy. 

\subsubsection{Utility of Randomized Response}

Suppose the real and the estimated counts of data owners' truthful ``yes'' answers are $\tilde{Y}$ and $Y_E$, respectively.  Then, the utility is defined as the relative error $\eta$ --- the magnitude of the difference between the real count and the estimated count, divided by the magnitude of the real count.
\begin{equation}
\label{eqn:eta}
\eta = \bigl|\frac{Y - Y_A }{Y}\bigr|
\end{equation}

Here, smaller relative error $\eta$ means higher utility of the randomized responses, and vice versa.

%% file: evaluation.tex
\section{Evaluation}

\begin{figure}[t!]
{\includegraphics[width=1\columnwidth]{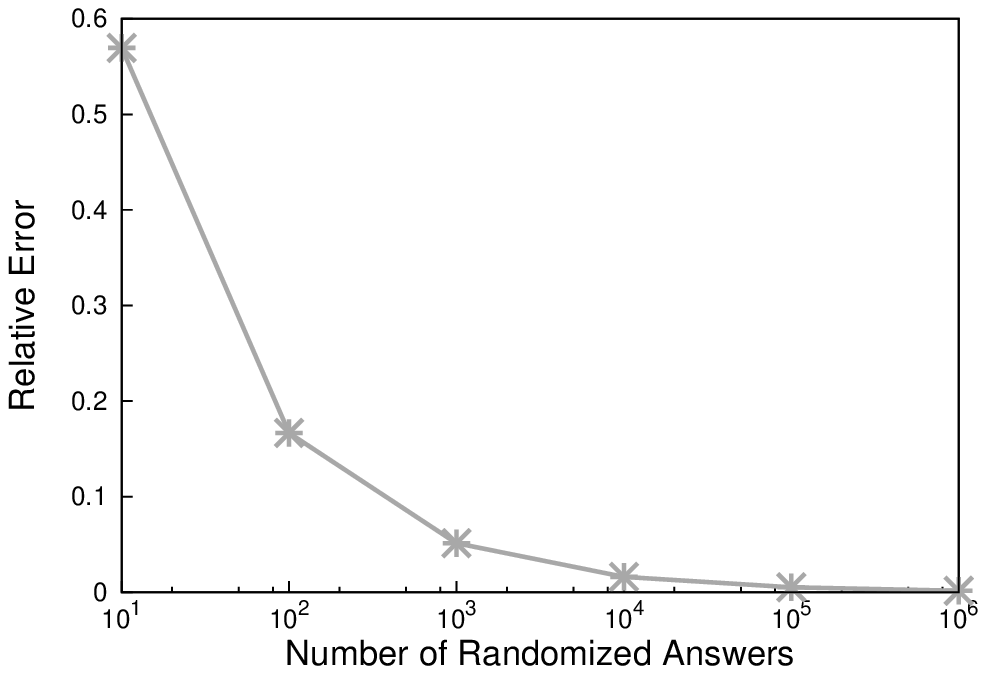}\label{fig:relativeerror-samplesize}}\hfill
{\includegraphics[width=1\columnwidth]{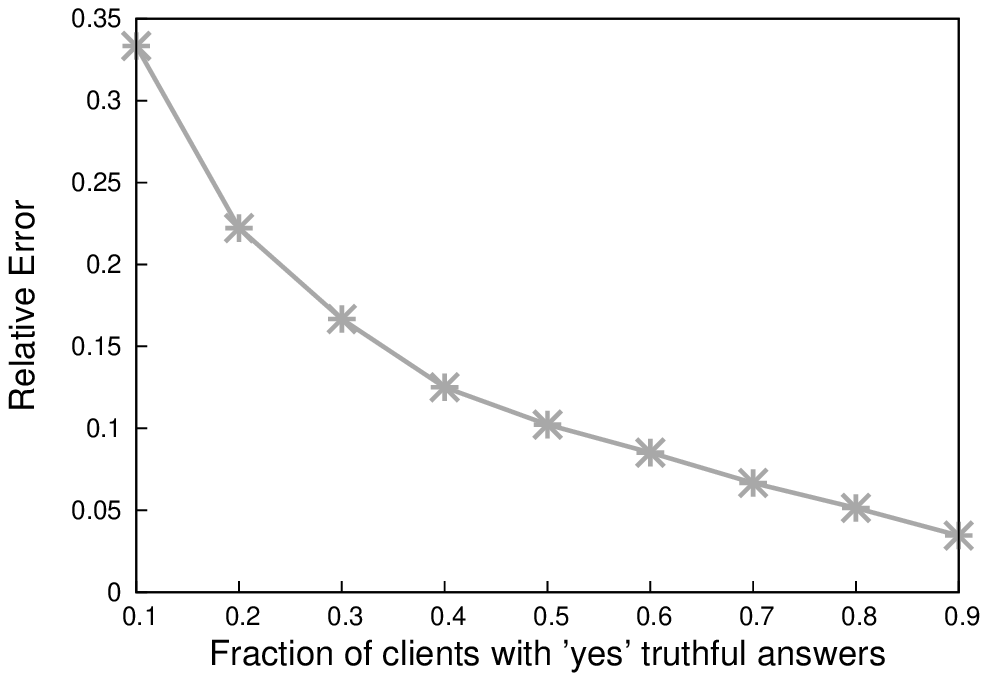}\label{fig:relativeerror-yesfraction}}\hfill
\protect\caption{The two coin flipping parameters $p$ and $q$ are set to achieve the highest possible utility.  (a) The fraction of the population with the sensitive attribute is 80\%.  (b) We also test with varying fractions having the sensitive attribute (different fractions of original ``yes'' answers), and the results follow the same pattern. The total number of data owners is 1000.}
\vspace{-2.0em}
\label{fig:relativeerrorresults}
\end{figure}

Privacy systems must carefully balance the trade-off between privacy and utility. Increasing privacy causes a corresponding decrease in utility and vice-versa. In distributed systems it becomes more difficult to balance this trade-off since there is no centralized system to add the minimum differentially private noise required.

The system evaluation involves varying the numbers and fractions
of data owners participating. In general, we seek to show that as more data owners participate the utility improves quickly due to the law of large numbers. This is particularly suitable for stream analytics whereby we can reasonably expect a large amount of data.

\begin{figure}[t!]
\centering
\includegraphics[width=.45\textwidth]{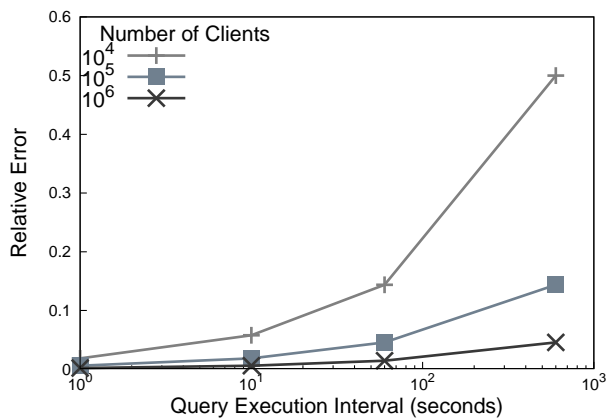}
\caption{Utility of query results. Here, the utility is measured by the query result's relative error.}
\vspace{-2.0em}
\label{fig:clientfrequency}
\end{figure}

Figure~\ref{fig:clientfrequency} shows that, in a typical large-scale distributed IoT application scenario with one million devices each answering a query every 10 seconds, the system can produce query results every one second with the relative error smaller than 0.5\%. As compared to prior systems~\cite{DBLP:conf/ndss/FriedmanSKS14,DBLP:conf/pet/ChanLSX12} we achieve both strong local privacy guarantees as well as high utility without requiring synchronization amongst system components or devices as each device independently privatizes their answer. The experimental results also shows that the utility improves with the increase of the number of devices, as well as with the decrease of the query execution interval.  This is because increasing the number of devices or reducing the query execution interval will put more answers into a batch.  As implied in Figure~\ref{fig:relativeerrorresults}, more answers in a batch lead to a lower relative error (i.e., higher utility).  In certain scenarios, if there are just a few devices or the devices answer queries less often, the system needs to extend the batching interval to put more answers into each batch for achieving high utility, although this may incur longer latency.

%% file: conclusion.tex
\section{Conclusion}

In this paper, a highly scalable, efficient, privacy preserving Differential Privacy scheme is introduced for use in future and emerging IoT and data stream analytics scenarios.  The scheme privatizes the data on the source device, thereby obviating many traditional attack and compromise modalities.  Moreover the computational complexity of the system is low, ensuring tractable implementation in both low-power IoT scenarios, as well as more capable aggregator nodes.  The system premise and functionality are validated through experimentation.

%% file: acknowledgements.tex
\section{Acknowledgements}

We would like to thank Ruichuan Chen for helpful discussions.